\documentclass[aps, prc, floatfix, nofootinbib, superscriptaddress, twocolumn]{revtex4-1}

\usepackage{latexsym}
\usepackage{amsmath}
\usepackage{amssymb}
\usepackage{amsfonts}

\usepackage[mathscr,scaled=1.15]{urwchancal}
\DeclareFontFamily{OT1}{pzc}{}
\DeclareFontShape{OT1}{pzc}{m}{it}%
{<-> s * [1.15] pzcmi7t}{}
\DeclareMathAlphabet{\mathpzc}{OT1}{pzc}{m}{it}

\usepackage{color}

\usepackage{supertabular}
\usepackage{placeins}
\usepackage{epsfig}
\usepackage{graphicx}
\usepackage{booktabs}

\definecolor{purple}{rgb}{0.5,0,0.5}
\definecolor{blue}{rgb}{0.0,0,0.9}
\definecolor{prdblue}{rgb}{0.133,0.118,0.498}
\usepackage[colorlinks=true, pdfstartview=FitV, linkcolor=prdblue, citecolor= prdblue, urlcolor=prdblue]{hyperref}

\begin{document}

\title{Charmonium mass shifts in an unquenched quark model}

\author{Xiaoyun Chen}
\email[]{xychen@jit.edu.cn} \affiliation{College of Science, Jinling Institute of Technology, Nanjing 211169, P. R.
China}

\author{Yue Tan}
\email[]{181001003@njnu.edu.cn}
\affiliation{Department of Physics, Yancheng Institute of Technology, Yancheng 224000, P. R. China}

\author{Youchang Yang}
\email[]{yangyc@gues.edu.cn}
\affiliation{Guizhou University of Engineering Science; Zunyi Normal University, P. R. China}

\begin{abstract}
In this paper, we performed a coupled-channel calculation and evaluated the mass shifts for all $1S$, $2S$, $1P$, $2P$ and $1D$ charmonium valence states below 4 GeV, by incorporating the four-quark components ($D$, $D^*$, $D_s$ and $D_s^*$ meson pairs) into the quark model. The valence-continuum coupling is provided by the $^3P_0$ quark-pair creation model. The induced mass shifts appear to be large and negative with the original transition operator in $^3P_0$ model, which raised up challenges for the valence quark model. More QCD-motivated models should be employed for the quark-pair creation Hamiltonian. So herein, we recalculated the mass shifts with the improved $^3P_0$ transition operator introduced in our previous work and the mass shifts are reduced by $75\%$ averagely. Besides, as a exercise, we adjust the confinement parameter $\Delta$ and recalculate the spectrum of the charmonium states. The masses of some charmonium states are reproduced well.
\end{abstract}

%\pacs{14.40.Be,12.39.Pn,24.10.Eq}

\maketitle

%%%%%%%%%%%%%%%%%%%%%%%%%%%%%%%%%%%%%%%%%%%%%%%%%%%%%%%%%%%%%%%%%%%%%%%%%%%%%%%%%%%%%%%%%%%%%%%%%%%%%%%%%%%%%%%%%%%%%%%

\section{Introduction} \label{introduction}
The discovery of many hidden charm states, the so-called $X, Y, Z$
mesons~\cite{epjc711534} and many bottomonium states, such as
$\eta_b(1S)$~\cite{Aubert:2008ba}, $
\Upsilon(^3D_J)$~\cite{delAmoSanchez:2010kz}
has created challenges for the conventional quenched quark model
and given great impetus to study on heavy quarkonium spectroscopy
recent years, because some members of them have unexpected
properties. Now it may be a good time to develop the unquenched quark model, in which, the effects of hadron loops (also called coupled-channel effects) were also considered.
%in the frame of the Cornell model \cite{Cornell}.
In recent years, the coupled-channel effects in the charmonium spectrum have been further studied ~\cite{prc77055206, prd76077502, prd72034010, prd80014012, Monteiro:2018rkg, plb7781,prd84034023,prd99014016}, and provided important information on the identifications of the newly reported states.

Godfrey and Isgur gave the predictions of the mass spectra of charmed and charmed-strange states in the nonrelativistic potential model~\cite{prd32189}. However, the observed masses are generally lower than the predicted ones, such as the narrow charm-strange mesons $D_{s_0}^*(2317)^+$~\cite{prl90242001} and $D_{s_1}(2460)^+$~\cite{prd68032002}, which also raised the special concern in both experiment and theory. The coupling to mesonic channels may be responsible for these anomalously low masses~\cite{prd68054006, prl91012003, prd70114013}.
$X(3872)$ is the most widely discussed state in the charmonium states. As the state sits just at the $D\bar{D}^*$ threshold, it
might be a $D\bar{D}^*$ molecule bound state. The study~\cite{Ortega:2009hj} indicated that it may be a mixture of a
$D\bar{D}^*$ molecule and the $\chi_{c_1}(2P)~(c\bar{c})$ considering the effects of coupled-channel using the $^3P_0$ quark-pair creation model. B. Q. Li \emph{et al.} supported the assignment of the $X(3872)$ as $\chi_{c_1}(2P)$-dominate charmonium state in two different
models: the coupled-channel model and the screened potential model~\cite{prd80014012}.  Recent study
by Zheng Cao and Qiang Zhao investigated the effects of $S$-wave
thresholds $D_{s_1}\bar{D}_s+c.c.$ and $D_{s_0}\bar{D}^*_s+c.c.$ on
vector charmonium spectrum, and found that it can lead to
formation of exotic states $Z_{cs}$ in the decay of $\psi(4415)
\rightarrow J/\psi K\bar{K}$~\cite{prd99014016}. There are many
other studies which presented good descriptions of the charmonium
states when considering the mass shifts induced by the
intermediate hadron loops~\cite{prd29110,prd76077502,pr429243}.

The $^3P_0$ model~\cite{npB10} is the simplest model for
light-quark pair creation which is widely applied in the effects
of hadron loops in the most of the above-mentioned papers. It assumes that the pair is created in the vacuum with the $^3P_0$ quantum numbers uniformly in space. The application of this model to the coupled-channels calculations has a long history. For example, T. Barnes has first reported results for hadronic mass shifts
of lower charmonium due to mixing with $D$, $D^*$, $D_s$ and
$D_s^*$ meson pairs, calculated within $^3P_0$ model and the
shifts appear to be alarmingly large~\cite{Barnes0412057}.
Refs.~\cite{prc77055206, prd411595,prd97094016} also arrived at
the same conclusion that $q\bar{q}~(q=u,~d)$ pairs were found to induce
very large mass shifts in the $^3P_0$ model. In our previous
work~\cite{prd97094016}, we compute the masses of ground state for the light mesons,
incorporating hadron loops in a chiral quark model using the
$^3P_0$ model to describe the pair creation, and explored the impact
of physically motivated modifications of the associated operator.
For the light-quark system, the coupling between $n\bar{n}~(n=u,~d)$ and meson
pairs component is weakened, producing mass shifts that are around
10\%$\sim$20\% of the hadron bare masses. In our present work, we will
keep exploring the effect of modified operator of the $^3P_0$
model for the heavy-quark charmonium system and trying to understand the
properties of the newly found charmonium states.

In this paper the effects of coupled-channels for charmonia levels
including all $1S$, $2S$, $1P$, $2P$ and $1D$ valence states
are presented. We calculated the mass shifts of these charmonium
states based on the nonrelativistic chiral quark model and solved
the quantum mechanics problem using the Gaussian expansion method
(GEM)~\cite{Hiyama:2003cu} instead of the simple harmonic
oscillator (SHO) ones~\cite{prc77055206, prd72034010, prd546811}. In
Sec. \ref{GEM and chiral quark model} the chiral quark model and the
GEM are outlined. Sec. \ref{sec3P0} introduces the
$^3P_0$ model briefly. And Sec. \ref{Numerical Results} is devoted to a
discussion of the results. In Sec. \ref{epilogue}, the paper ends
with a short summary.

%%%%%%%%%%%%%%%%%%%%%%%%%%%%%%%%%%%%%%%%%%%%%%%%%%%%%%%%%%%%%%%%%%%%%%%%%%%%%%%%%%%%%%%%%%%%%%%%%%%%%%%%%%%%%%%%%%%%%%%

\section{Chiral quark model}
\label{GEM and chiral quark model}
In the nonrelativistic quark model, we obtained the meson spectrum by solving
a Schr\"{o}dinger equation:
\begin{equation}
\label{Hamiltonian1} H \Psi_{M_I M_J}^{IJ} (1,2) =E^{IJ} \Psi_{M_I
M_J}^{IJ} (1,2)\,,
\end{equation}
where $1$, $2$ represents the quark and antiquark labels. $\Psi_{M_I M_J}^{IJ}(1,2)$ is the wave function of a meson composed of a quark and a antiquark with quantum numbers $IJ^{PC}$ and reads,
\begin{align}
\nonumber
& \Psi_{M_I M_J}^{IJ}(1,2) \\
& =\sum_{\alpha}C_{\alpha} \left[
\psi_{l}(\mathbf{r})\chi_{s}(1,2)\right]^{J}_{M_J}
\omega^c(1,2)\phi^I_{M_I}(1,2), \label{PsiIJM}
\end{align}
where $\psi_{l}(\mathbf{r})$, $\chi_{s}(1,2)$,
$\omega^c(1,2)$, $\phi^I(1,2)$ are orbit, spin, color and flavor wave
functions, respectively. $\alpha$ denotes the intermediate quantum numbers, $l,s$ and
possible flavor indices. In our calculations, the orbital wave functions is expanded using a series of Gaussians,
\begin{subequations}
\label{radialpart}
\begin{align}
\psi_{lm}(\mathbf{r}) & = \sum_{n=1}^{n_{\rm max}} c_{n}\psi^G_{nlm}(\mathbf{r}),\\
\psi^G_{nlm}(\mathbf{r}) & = N_{nl}r^{l}
e^{-\nu_{n}r^2}Y_{lm}(\hat{\mathbf{r}}),
\end{align}
\end{subequations}
with the Gaussian size parameters chosen according to the
following geometric progression
\begin{equation}\label{gaussiansize}
\nu_{n}=\frac{1}{r^2_n}, \quad r_n=r_1a^{n-1}, \quad
a=\left(\frac{r_{n_{\rm max}}}{r_1}\right)^{\frac{1}{n_{\rm
max}-1}}.
\end{equation}
This procedure enables optimization of the ranges using just a
small number of Gaussians.

At this point, the wave function in Eq.\,\eqref{PsiIJM} is expressed as follows:
\begin{align}
\nonumber
&\Psi_{M_I M_J}^{IJ}(1,2) \\
& =\sum_{n\alpha} C_{\alpha}c_n
 \left[ \psi^G_{nl}(\mathbf{r})\chi_{s}(1,2) \right]^{J}_{M_J}\omega^c(1,2)\phi^I_{M_I}(1,2).\label{Gauss1}
\end{align}
We employ the Rayleigh-Ritz variational principle for solving the
Schr\"{o}dinger equation due to the non-orthogonality of Gaussians, which leads to a generalized eigenvalue problem
\begin{subequations}
\label{HEproblem}
\begin{align}
 \sum_{n^{\prime},\alpha^{\prime}} & (H_{n\alpha,n^{\prime}\alpha^{\prime}}^{IJ}
-E^{IJ} N_{n\alpha,n^{\prime}\alpha^{\prime}}^{IJ}) C_{n^{\prime}\alpha^{\prime}}^{IJ} = 0, \\
 &H_{n\alpha,n^{\prime}\alpha^{\prime}}^{IJ} =
  \langle\Phi^{IJ}_{M_I M_J,n\alpha}| H | \Phi^{IJ}_{M_I M_J,n^{\prime}\alpha^{\prime}}\rangle ,\\
 &N_{n\alpha,n^{\prime}\alpha^{\prime}}^{IJ}=
  \langle\Phi^{IJ}_{M_I M_J,n\alpha}|1| \Phi^{IJ}_{M_I M_J,n^{\prime}\alpha^{\prime}}\rangle,
\end{align}
\end{subequations}
with
$\Phi^{IJ}_{M_I M_J,n\alpha} =
[\psi^G_{nl}(\mathbf{r})\chi_{s}(1,2) ]^{J}_{M_J}
\omega^c(1,2)\phi^I_{M_I}(1,2)$,
$C_{n\alpha}^{IJ} = C_{\alpha}c_n$.

We get the mass of the four-quark system also by solving
a Schr\"{o}dinger equation:
\begin{equation}
    H \, \Psi^{4\,IJ}_{M_IM_J}=E^{IJ} \Psi^{4\,IJ}_{M_IM_J},
\end{equation}
where $\Psi^{4\,IJ}_{M_IM_J}$ is the wave function of the
four-quark system, which can be constructed as follows. In our calculations, we only consider the meson-meson picture with the color singlet for the four quark system in coupled-channel effects. First, we write down the wave functions of two meson clusters,
\begin{subequations}
\begin{align}
\nonumber
&    \Psi^{I_1J_1}_{M_{I_1}M_{J_1}}(1,2)=\sum_{\alpha_1 n_1} {\mathpzc C}^{\alpha_1}_{n_1} \\
    & \times  \left[ \psi^G_{n_1 l_1}(\mathbf{r}_{12})\chi_{s_1}(1,2)\right]^{J_1}_{M_{J_1}}
 \omega^{c_1}(1,2)\phi^{I_1}_{M_{I_1}}(1,2),   \\
&    \Psi^{I_2J_2}_{M_{I_2}M_{J_2}}(3,4)=\sum_{\alpha_2 n_2} {\mathpzc C}^{\alpha_2}_{n_2} \nonumber \\
    & \times \left[ \psi^G_{n_2 l_2}(\mathbf{r}_{34})\chi_{s_2}(3,4)\right]^{J_2}_{M_{J_2}}
    \omega^{c_2}(3,4)\phi^{I_2}_{M_{I_2}}(3,4),
\end{align}
\end{subequations}
then the total wave function of the four-quark state is:
\begin{align}
& \Psi^{4\,IJ}_{M_IM_J}  =  {\cal A} \sum_{L_r}\left[
\Psi^{I_1J_1}(1,2)\Psi^{I_2J_2}(3,4)
     \psi_{L_r}(\mathbf{r}_{1234})\right]^{IJ}_{M_IM_J}    \nonumber \\
\nonumber
  & =  \sum_{\alpha_1\,\alpha_2\,n_1\,n_2\,L_r}
  {\mathpzc C}^{\alpha_1}_{n_1} {\mathpzc C}^{\alpha_2}_{n_2} \bigg[ \left[\psi^G_{n_1 l_1}(\mathbf{r}_{12})\chi_{s_1}(1,2)\right]^{J_1} \nonumber \\
& \quad \times
            \left[\psi^G_{n_2 l_2}(\mathbf{r}_{34})\chi_{s_2}(3,4)\right]^{J_2}
             \psi_{L_r}(\mathbf{r}_{1234})\bigg]^{J}_{M_J} \nonumber \\
 &      \quad  \times \left[\omega^{c_1}(1,2)\omega^{c_2}(3,4)\right]^{[1]}
     \left[\phi^{I_1}(1,2)\phi^{I_2}(3,4)\right]^{I}_{M_I},
\end{align}
Here, ${\cal A}$ is the antisymmetrization operator: if all quarks
(antiquarks) are taken as identical particles, then
\begin{equation}
{\cal A}=\frac{1}{2}(1-P_{13}-P_{24}+P_{13}P_{24}).
\end{equation}
$\psi_{L_r}(\mathbf{r}_{1234})$ is the two-cluster relative
wave function which is also expanded in a series of Gaussians.
$L_r$ describes the relative cluster orbital angular momentum. Need to be noted that, in our calculations, the angular momentum for the two mesons $l_1$ and $l_2$ equals zero. So for the $1S$, $2S$ and $1D$ states, the relative angular momentum $L_r$ equals 1 ($P$ wave); for the $1P$ and $2P$ states, we only consider the $L_r=0$ with $S$ wave between the two clusters, and $L_r=2$ with $D$ wave is not considered herein, which is our future work.
For the quark model introduction, we take four-quark system as an example. (The two-quark system is relative simple, here we will omit it). The Hamiltonian of the chiral quark model for the four-quark system consists of three parts:
quark rest mass, kinetic energy, potential energy:
\begin{align}
 H & = \sum_{i=1}^4 m_i  +\frac{p_{12}^2}{2\mu_{12}}+\frac{p_{34}^2}{2\mu_{34}}
  +\frac{p_{r}^2}{2\mu_{r}}  \quad  \nonumber \\
  & + \sum_{i<j=1}^4 \left( V_{\rm CON}^{C}(\boldsymbol{r}_{ij})+ V_{\rm OGE}^{C}(\boldsymbol{r}_{ij}) \right. \quad  \nonumber \\
  & \left. + V_{\rm CON}^{SO}(\boldsymbol{r}_{ij}) + V_{\rm OGE}^{SO}(\boldsymbol{r}_{ij}) +\sum_{\chi=\pi,K,\eta} V_{ij}^{\chi}
   +V_{ij}^{\sigma}\right).
\end{align}
Where $m_i$ is the constituent mass of $i$th quark (antiquark). $\frac{\bf{p^2_{ij}}}{2\mu_{ij}} (ij=12; 34)$ and $\frac{\bf{p^2_{r}}}{2\mu_{r}}$ represents the inner kinetic of two-cluster and the relative motion kinetic between two clusters, respectively, with
\begin{subequations}
\begin{align}
\bf{p}_{12}&=\frac{m_2\mathbf{p}_1-m_1\mathbf{p}_2}{m_1+m_2}, \\
\mathbf{p}_{34}&=\frac{m_4\mathbf{p}_3-m_3\mathbf{p}_4}{m_3+m_4},  \\
\mathbf{p}_{r}&= \frac{(m_3+m_4)\mathbf{p}_{12}-(m_1+m_2)\mathbf{p}_{34}}{m_1+m_2+m_3+m_4}, \\
\mu_{ij}&=\frac{m_im_j}{m_i+m_j}, \\
\mu_{r}&=\frac{(m_1+m_2)(m_3+m_4)}{m_1+m_2+m_3+m_4}.
\end{align}
\end{subequations}
$V_{\rm CON}^{C}$ and $V_{\rm OGE}^{C}$ is the central part of the confinement and central part of one-gluon-exchange. $V_{\rm CON}^{SO}$ and $V_{\rm OGE}^{SO}$ is the noncentral potential energy. $V_{ij}^{\chi=\pi, K, \eta}$, and $\sigma$ exchange represents the one Goldstone boson exchange. The forms of the potentials are \cite{Valcarce:2005em}: {\allowdisplaybreaks
\begin{subequations}
\begin{align}
V_{\rm CON}^{C}(\boldsymbol{r}_{ij})&= ( -a_c r_{ij}^2-\Delta ) \boldsymbol{\lambda}_i^c \cdot \boldsymbol{\lambda}_j^c ,  \\
V_{\rm CON}^{\rm SO}(\boldsymbol{r}_{ij})&=\boldsymbol{\lambda}_i^c \cdot \boldsymbol{\lambda}_j^c \cdot \frac{-a_c}{2m_i^2m_j^2}\bigg\{ \bigg((m_i^2+m_j^2)(1-2a_s) \nonumber \\
&+4m_im_j(1-a_s)\bigg)(\boldsymbol{S_{+}} \cdot \boldsymbol{L})+(m_j^2-m_i^2) \nonumber \\
&(1-2a_s)(\boldsymbol{S_{-}} \cdot \boldsymbol{L})\bigg\}, \\
 V_{\rm OGE}^{C}(\boldsymbol{r}_{ij})&= \frac{\alpha_s}{4} \boldsymbol{\lambda}_i^c \cdot \boldsymbol{\lambda}_{j}^c
\left[\frac{1}{r_{ij}}-\frac{2\pi}{3m_im_j}\boldsymbol{\sigma}_i\cdot
\boldsymbol{\sigma}_j
  \delta(\boldsymbol{r}_{ij})\right],  \\
V_{\rm OGE}^{\rm SO}(\boldsymbol{r}_{ij})&=-\frac{1}{16}\cdot\frac{\alpha_s}{m_i^2m_j^2}\boldsymbol{\lambda}_i^c
\cdot \boldsymbol{\lambda}_j^c\big\{\frac{1}{r_{ij}^3}-\frac{e^{-r_{ij}/r_g(\mu)}}{r_{ij}^3}\cdot \nonumber \\
&(1+\frac{r_{ij}}{r_g(\mu)})\big\}\times \bigg\{\bigg((m_i+m_j)^2+2m_im_j\bigg)\nonumber \\
&(\boldsymbol{S_{+}} \cdot \boldsymbol{L})+(m_j^2-m_i^2)(\boldsymbol{S_{-}} \cdot \boldsymbol{L})\bigg\},  \\
\delta{(\boldsymbol{r}_{ij})} & =  \frac{e^{-r_{ij}/r_0(\mu_{ij})}}{4\pi r_{ij}r_0^2(\mu_{ij})}, \mathbf{S}_{\pm}=\mathbf{S}_1\pm \mathbf{S}_2,\\
V_{\pi}(\boldsymbol{r}_{ij})&= \frac{g_{ch}^2}{4\pi}\frac{m_{\pi}^2}{12m_im_j}
  \frac{\Lambda_{\pi}^2}{\Lambda_{\pi}^2-m_{\pi}^2}m_\pi v_{ij}^{\pi}
  \sum_{a=1}^3 \lambda_i^a \lambda_j^a,  \\
V_{K}(\boldsymbol{r}_{ij})&= \frac{g_{ch}^2}{4\pi}\frac{m_{K}^2}{12m_im_j}
  \frac{\Lambda_K^2}{\Lambda_K^2-m_{K}^2}m_K v_{ij}^{K}
  \sum_{a=4}^7 \lambda_i^a \lambda_j^a,   \\
\nonumber V_{\eta} (\boldsymbol{r}_{ij})& =
\frac{g_{ch}^2}{4\pi}\frac{m_{\eta}^2}{12m_im_j}
\frac{\Lambda_{\eta}^2}{\Lambda_{\eta}^2-m_{\eta}^2}m_{\eta}
v_{ij}^{\eta}  \\
 & \quad \times \left[\lambda_i^8 \lambda_j^8 \cos\theta_P
 - \lambda_i^0 \lambda_j^0 \sin \theta_P \right],   \\
v_{ij}^{\chi}(\boldsymbol{r}_{ij}) & =  \left[ Y(m_\chi r_{ij})-
\frac{\Lambda_{\chi}^3}{m_{\chi}^3}Y(\Lambda_{\chi} r_{ij})
\right]
\boldsymbol{\sigma}_i \cdot\boldsymbol{\sigma}_j,\\
V_{\sigma}(\boldsymbol{r}_{ij})&= -\frac{g_{ch}^2}{4\pi}
\frac{\Lambda_{\sigma}^2}{\Lambda_{\sigma}^2-m_{\sigma}^2}m_\sigma \nonumber \\
& \quad \times \left[
 Y(m_\sigma r_{ij})-\frac{\Lambda_{\sigma}}{m_\sigma}Y(\Lambda_{\sigma} r_{ij})\right]  ,
\end{align}
\end{subequations}}
\hspace*{-0.5\parindent}%
where $\mathbf{S}_1$ and $\mathbf{S}_2$ is the spin of the two meson clusters. $Y(x)  =   e^{-x}/x$; $r_0(\mu_{ij}) =s_0/\mu_{ij}$; $\boldsymbol{\sigma}$ are the $SU(2)$ Pauli matrices; $\boldsymbol{\lambda}$, $\boldsymbol{\lambda}^c$ are $SU(3)$ flavor, color Gell-Mann matrices, respectively; $g^2_{ch}/4\pi$ is the chiral coupling constant, determined from the $\pi$-nucleon coupling; and $\alpha_s$ is an effective scale-dependent running coupling
\cite{Valcarce:2005em},
\begin{equation} \label{alphas}
\alpha_s(\mu_{ij})=\frac{\alpha_0}{\ln\left[(\mu_{ij}^2+\mu_0^2)/\Lambda_0^2\right]}.
\end{equation}
In our calculations, for the two-quark system, besides the central potential energy, the noncentral potential energy is also included. But in the four-quark system calculations, we find that the influence of the noncentral potential energy on the mass shift of the state is tiny.

Lastly, we show the model parameters \cite{Vijandemodel} in Table \ref{modelparameters}. Need to be noted that, in the reference \cite{Vijandemodel}, the confinement item takes the form $V^{C}_{ij}=\big(-a_c(1-e^{-\mu_c r_{ij}}\big)+\Delta)(\boldsymbol{\lambda}_i^c \cdot \boldsymbol{\lambda}_j^c)$. And in our present calculations, the usual quadratic confinement $V^{C}_{ij}=( -a_c r_{ij}^2-\Delta ) \boldsymbol{\lambda}_i^c \cdot \boldsymbol{\lambda}_j^c$ is employed, so some parameters are different such as quark mass, $a_c$ and $\Delta$.

Using the model parameters, we calculated the masses of some mesons from light to heavy, especially the relevant charmonium $c\bar{c}$ mesons $\eta_c$, $J/\psi$, $\chi_{c_J} (J=0,1,2)$, $h_c$ in the chiral quark model, which are demonstrated in Table \ref{mesonspectra}. In order to obtain the stable masses, we take the gaussian size parameters $r_1=0.01$, $r_n=2$, $n=16$ in Eq. (\ref{gaussiansize}). From the table, we can find that the quark model achieves great success on describing the hadron spectra, especially for the ground-state mesons such as most light mesons and heavy mesons $\eta_c(1S), J/\psi(1S)$. But it still be faced some challenges on the charmonium excited states such as $\eta_c(2S), \psi(2S), \chi_{c_J}(1P)$, $\chi_{c_J}(2P)$ and $1D$ states since more higher charmonium states have been observed experimentally. For $b\bar{b}$ system, the masses of the ground-state $\eta_b(1S)$ and $\Upsilon(1S)$ are not so satisfactory, but for the excited states, the masses are well consistent with the experimental values unexpectedly such as $\Upsilon(2S), \chi_{b_J}(1P)$ and $\chi_{b_J}(2P)$.

\begin{table}[!t]
\begin{center}
\caption{ \label{modelparameters} Model parameters, determined by
fitting the meson spectrum, leaving room for unquenching
contributions in the case of light-quark systems.}
\begin{tabular}{llr}
\hline\noalign{\smallskip}
Quark masses   &$m_u=m_d$     &313  \\
   (MeV)       &$m_s$         &536  \\
               &$m_c$         &1728 \\
               &$m_b$         &5112 \\
\hline
Goldstone bosons                &$m_{\pi}$        &0.70  \\
(fm$^{-1} \sim 200\,$MeV )      &$m_{\sigma}$     &3.42  \\
                                &$m_{\eta}$       &2.77  \\
                                &$m_{K}$          &2.51  \\
                                &$\Lambda_{\pi}=\Lambda_{\sigma}$     &4.2  \\
                                &$\Lambda_{\eta}=\Lambda_{K}$         &5.2  \\
                   \cline{2-3}
                               &$g_{ch}^2/(4\pi)$                     &0.54  \\
                               &$\theta_p(^\circ)$                    &-15 \\
\hline
Confinement                    &$a_c$ (MeV fm$^{-2}$)                 &101 \\
                               &$\Delta$ (MeV)                        &-78.3 \\
\hline
OGE                            & $\alpha_0$                           &3.67 \\
                               &$\Lambda_0({\rm fm}^{-1})$            &0.033 \\
                               &$\mu_0$(MeV)                          &36.98 \\
                               &$s_0$(MeV)                            &28.17 \\
\hline
\end{tabular}
\end{center}
\end{table}

\linespread{1.2}
\begin{table}[!t]
\begin{center}
\renewcommand\tabcolsep{10.0pt} % 调整表格列间的宽度
\caption{ \label{mesonspectra} The mass spectrum in the chiral quark model, in comparison with the experimental data \cite{PDG} (in unit of MeV).}
\begin{tabular}{cccc}
\hline\hline\noalign{\smallskip}
Name             &$J^{P(C)}$           &Mass     &PDG \cite{PDG}\\
\hline
$\pi$             &$0^-$                &134.9    &135.0 \\
$K$               &$0^-$                &489.4    &493.7 \\
$\rho$            &$1^{--}$             &772.3    &775.3 \\
$K^*$             &$1^-$                &913.6    &892.0\\
$\omega$          &$1^{--}$             &701.6    &782.7 \\
$\eta$            &$0^{-+}$             &669.2    &547.9\\
$\phi(1020)$      &$1^{--}$             &1015.9   &1019.5 \\
$D^0$             &$0^-$                &1861.9   &1864.8 \\
$D^{*0}$          &$1^-$                &1980.6   &2006.9\\
$D_s^+$           &$0^-$                &1950.1   &1968.4 \\
$D_s^{*+}$        &$1^-$                &2079.9   &2112.2\\
$B^-$             &$0^-$                &5280.7   &5279.3 \\
$B^{*}$           &$1^-$                &5319.6   &5324.7\\
$B_s^0$           &$0^-$                &5367.4   &5366.9 \\
$B_s^*$           &$1^-$                &5410.2   &5415.4\\
$\eta_c(1S)$      &$0^{-+}$             &2964.4   &2983.9\\
$\eta_c(2S)$      &$0^{-+}$             &3507.8   &3637.5\\
$J/\psi$          &$1^{--}$             &3096.4   &3096.0\\
$\psi(2S)$        &$1^{--}$             &3605.0   &3686.1\\
$\chi_{c_0}(1P)$  &$0^{++}$             &3362.8   &3414.7\\
$\chi_{c_0}(2P)$  &$0^{++}$             &3814.7   &$\chi_{c_0}(3915)?$\\
$\chi_{c_1}(1P)$  &$1^{++}$             &3393.9   &3510.7\\
$\chi_{c_1}(2P)$  &$1^{++}$             &3851.9   &$\chi_{c_1}(3872)?$\\
$\chi_{c_2}(1P)$  &$2^{++}$             &3435.8   &3556.2\\
$\chi_{c_2}(2P)$  &$2^{++}$             &3901.1   &$\chi_{c_2}(3930)?$\\
$h_c(1P)$         &$1^{+-}$             &3416.1   &3525.4\\
$h_c(2P)$         &$1^{+-}$             &3877.4   &$Z_c(3900)$?\\
$\eta_b(1S)$      &$0^{-+}$             &9561.5   &9398.7 \\
$\Upsilon(1S)$    &$1^{--}$             &9647.8   &9460.3\\
$\Upsilon(2S)$    &$1^{--}$             &10016.7  &10023.3\\
$\chi_{b_0}(1P)$  &$0^{++}$             &9916.8   &9859.4\\
$\chi_{b_0}(2P)$  &$0^{++}$             &10198.4  &10232.5\\
$\chi_{b_1}(1P)$  &$1^{++}$             &9925.4   &9892.8\\
$\chi_{b_1}(2P)$  &$1^{++}$             &10208.2  &10255.5\\
$\chi_{b_2}(1P)$  &$2^{++}$             &9938.9   &9912.2\\
$\chi_{b_2}(2P)$  &$2^{++}$             &10223.2  &10268.7\\
$h_b(1P)$         &$1^{+-}$             &9932.4   &9899.3\\
$h_b(2P)$         &$1^{+-}$             &10216.1  &10259.8\\
$\eta_{c_2}(1D)$  &$2^{-+}$             &3675.1   &? \\
$\psi(1D)$        &$1^{--}$             &3653.3   &$\psi(3770)$? \\
$\psi_2(1D)$      &$2^{--}$             &3668.3   &$\psi_2(3823)$?\\
$\psi_3(1D)$      &$3^{--}$             &3688.1   &$\psi_3(3842)$?\\
\hline\hline
\end{tabular}
\end{center}
\end{table}

%%%%%%%%%%%%%%%%%%%%%%%%%%%%%%%%%%%%%%%%%%%%%%%%%%%%%%%%%%%%%%%%%%%%%%%%%%%%%%%%%%%%%%%%%%%%%%%%%%%%%%%%%%%%%%%%%%%%%%%

\section{$^3P_0$ model}
\label{sec3P0}
The $^3P_0$ quark-pair creation model \cite{npB10,
LeYaouanc:1972vsx, LeYaouanc:1973ldf} has been widely applied to
OZI rule allowed two-body strong decays of hadrons
\cite{Capstick:1986bm, Roberts:1992js, Capstick:1993kb,
Page:1995rh, Ackleh:1996yt, Segovia:2012cd}. If the quark and antiquark in the source meson are labeled by 1, 2, and the quark and antiquark ($u\bar{u}$, $d\bar{d}$, $s\bar{s}$) generated in the vacuum are numbered as 3, 4, the operator of the $^3P_0$ model reads:
\begin{align} \label{T0}
T_0 & =-3\, \gamma \sum_m\langle 1m1(-m)|00\rangle\int
d\mathbf{p}_3d\mathbf{p}_4\delta^3(\mathbf{p}_3+\mathbf{p}_4)\nonumber\\
& \quad \times{\cal{Y}}^m_1(\frac{\mathbf{p}_3-\mathbf{p}_4}{2})
\chi^{34}_{1-m}\phi^{34}_0\omega^{34}_0b^\dagger_2(\mathbf{p}_3)d^\dagger_3(\mathbf{p}_4),
\end{align}
where $\gamma$ describes the probability for creating a
quark-antiquark pair with momenta $\mathbf{p}_3$ and
$\mathbf{p}_4$ from the $0^{++}$ vacuum. It is normally determined by fitting an array of hadron strong decays. This yields $\gamma=6.95$ for $u\bar{u}$ and $d\bar{d}$ pair creation, and $\gamma=6.95/\sqrt{3}$ for $s\bar{s}$ pair creation \cite{LeYaouanc:1977gm}. $\omega^{34}_0$ and $\phi^{34}_{0}$ are the color and flavor wave function components, respectively.

\section{Numerical Results}
\label{Numerical Results}
By incorporating the four-quark components ($c\bar{c}q\bar{q}~(q=u,d,s)$) into the charmonium
$c\bar{c}$ mesons, we can get the eigenvalues of the $c\bar{c}+c\bar{c}q\bar{q}$ system by solving the Schr\"{o}dinger
equation,
\begin{eqnarray}
H\Psi=E\Psi ,
\end{eqnarray}
where $\Psi$ and $H$ is the wave function and the Hamiltonian of
the system, it takes,
\begin{eqnarray}
\Psi=c_1\Psi_{2q}+c_2\Psi_{4q} \,,\\
 H=H_{2q}+H_{4q}+T_{0}\,.
\end{eqnarray}
Because the number of particles is conserved in the
nonrelativistic quark model, the $H_{2q}$ only acts on the wave
function of two-quark $c\bar{c}$ system, $\Psi_{2q}$, and the $H_{4q}$ only acts on the wave function of four-quark system, $\Psi_{4q}$. The transition operator $T_0$ (Eq. (\ref{T0})) in the $^3P_0$ model is responsible for the coupling of the two- and four-quark system.

In this way, we can get the matrix elements of the Hamiltonian,
\begin{align}
\langle\Psi| & H|\Psi\rangle = \langle
c_1\Psi_{2q}+c_2\Psi_{4q}|H|c_1\Psi_{2q}+c_2\Psi_{4q}\rangle
\nonumber \\
&=c_1^2\langle\Psi_{2q}|H_{2q}|\Psi_{2q}\rangle+c_2^2\langle\Psi_{4q}|H_{4q}|\Psi_{4q}\rangle
\nonumber \\
&\quad+c_1c_2^*\langle\Psi_{4q}|T_{0}|\Psi_{2q}\rangle+c_1^*c_2\langle\Psi_{2q}|T_{0}^{\dagger}|\Psi_{4q}\rangle,
\end{align}
and the block-matrix structure for the Hamiltonian and overlap
takes,
\begin{equation}
(H)=\left[\begin{array}{cc} (H_{2q}) & (H_{24})\\
(H_{42}) & (H_{4q})
\end{array}
\right],
(N)=\left[\begin{array}{cc} (N_{2q}) & (0)\\
(0) & (N_{4q})
\end{array}
\right] \,,
\end{equation}
with
{\allowdisplaybreaks
\begin{subequations}
\begin{align}
 (H_{2q})&=\langle\Psi_{2q}|H_{2q}|\Psi_{2q}\rangle, \\
 (H_{24})&=\langle\Psi_{4q}|T_{0}|\Psi_{2q}\rangle, \\
 (H_{4q})&=\langle\Psi_{4q}|H_{4q}|\Psi_{4q}\rangle,\\
(N_{2q})&=\langle\Psi_{2q}|1|\Psi_{2q}\rangle, \\
(N_{4q})&=\langle\Psi_{4q}|1|\Psi_{4q}\rangle.
\end{align}
\end{subequations}
}
Where $(H_{2q})$ and $(H_{4q})$ is the matrix for the pure two-quark $c\bar{c}$
system and pure four-quark system, respectively. $(H_{24})$ is the coupling matrix of two-quark system and four-quark system.

Finally the eigenvalues ($E_n$) and eigenvectors ($C_n$) of the system are obtained by solving the diagonalization problem,
\begin{eqnarray}
\Big[
\begin{array}{c}
(H)-E_n(N)
\end{array}
\Big] \Big[
\begin{array}{c} C_n
\end{array}
\Big]=0. \label{geig}
\end{eqnarray}

In our calculations, a convergence factor  $e^{-f^2 \mathbf{p}^2}$ was inserted into
the operator $T_0$ in Eq.~(\ref{T0}) in order to be Fourier
transformed, because the two- and four-quark system are solved in
coordinate space. The Fourier transformed factor is written as,
\begin{eqnarray}\label{Tr}
T_1&=&-i3\gamma\sum_{m}\langle 1m1(-m)|00\rangle\int
d\mathbf{r_3}d\mathbf{r_4}(\frac{1}{2\pi})^{\frac{3}{2}}2^{-\frac{5}{2}}f^{-5} \nonumber \\
 &&rY_{1m}(\hat{\mathbf{r}})
 {\rm e}^{-\frac{\mathbf{r}^2}{4f^2}}\chi_{1-m}^{34}\omega_{0}^{34} \phi_{0}^{34}
 b_3^{\dagger}(\mathbf{r_3})d_4^{\dagger}(\mathbf{r_4}) .
\end{eqnarray}
There is one more parameter $f$ in the transition operator $T_1$.
When $f$ takes the limit to zero, the original form of the $^3P_0$
quark model is recovered. By the way, $\mathbf{r}$ in
Eq.~(\ref{Tr}) is the relative distance between the quark pair in
the vacuum, $\mathbf{r} = \mathbf{r}_3 - \mathbf{r}_4$.

By solving Eq.~(\ref{geig}) with the transition operator in Eq.~(\ref{Tr}) and in the limit $f \rightarrow 0$, $\gamma=6.95$, we obtained the
mass shifts for $\eta_c(1S)$, $\eta_c(2S)$, $J/\psi(1S)$, $J/\psi(2S)$, $\chi_{c_0}(1P)$, $\chi_{c_1}(1P)$, $\chi_{c_2}(1P)$, $h_c(1P)$ charmonium valence states, as well as the higher charmoium $2P$ and $1D$ states, by incorporating the four-quark components ($D$, $D^*$, $D_s$ and $D_s^*$ meson pairs) into the two-quark $c\bar{c}$ system. The results are shown in Table ~\ref{result1}. 
In order to get the stable mass shifts of the states, we take the gaussian size parameters $r_1=0.01$, $r_n=2$, $n=16$ in Eq. (\ref{gaussiansize}) for the two-quark charmonium system. For the four-quark system, we take $r_1=0.1$, $r_n=2$, $n=8$ for inner two meson pairs, and the relative gaussian size parameters between the two meson pairs take $r_1=0.1$, $r_n=6$, $n=9$.

There exist three open channels in our calculations, $\chi_{c_0}(2P) \rightarrow D\bar{D}$, $\chi_{c_1}(2P) \rightarrow D\bar{D}^*$, $h_c(2P) \rightarrow D\bar{D}^*$. For these open channels, the mass shifts of the states will change with the Gaussian distribution. Especially, the mass shifts will change with the increasing of spatial volume periodically. In our calculations, we picked up the biggest mass shifts as the contributions of this open channel by varying the Gaussian size parameter $r_n$ between the two meson pairs. 
Let us take the channel $\chi_{c_0}(2P) \rightarrow D\bar{D}$ as an example. For $D\bar{D}$ state, it has the discrete energy levels which will change with the varying Gaussian distribution in the theoretical calculations even if it is a scattering state. When considering the coupling of the $D\bar{D}$ and $\chi_{c_0}(2P)$, the strength of coupling will be increased as the energy of $D\bar{D}$ state is close to that of $\chi_{c_0}(2P)$, and the induced mass shift will become bigger. We take the biggest one as the mass shift of the state $\chi_{c_0}(2P)$ to the $D\bar{D}$ state.
Besides, if we expand the space further with higher $r_n$ values, the same biggest mass shift will be repeated. From the table, we can also find that for the open channels, the mass shifts are always larger than the close channels.

In Table \ref{result1}, the bare mass of the states are obtained in the quenched quark model, viz. solved with only the $c\bar{c}$ component. When considering the coupled-channel effects, we get the large negative mass shifts. Such large shifts invalidate the traditional quenched quark model.
In our previous work~\cite{prd97094016}, when we investigate the hadron loop effects of the $n\bar{n}~(n=u, d)$ states, similarly, large mass shifts are obtained with the original operator $T_1$ in Eq. (\ref{Tr}) in the limit $f \rightarrow 0$. In order to develop a more realistic unquenching procedure, in the work~\cite{prd97094016}, we gave some modifications of the operator $T_1$ . It reads,
\begin{align} \label{T2}
T_2&= -3\gamma\sum_{m}\langle 1m1(-m)|00\rangle\int
d\mathbf{r_3}d\mathbf{r_4}(\frac{1}{2\pi})^{\frac{3}{2}}ir2^{-\frac{5}{2}}f^{-5}
\nonumber \\
 & Y_{1m}(\hat{\mathbf{r}})
 {\rm e}^{-\frac{\mathbf{r}^2}{4f^2}}
 {\rm e}^{-\frac{R_{AV}^2}{R_0^2}}\chi_{1-m}^{34}\phi_{0}^{34}
 \omega_{0}^{34}b_3^{\dagger}(\mathbf{r_3})d_4^{\dagger}(\mathbf{r_4}),
\end{align}
compared with Eq.~(\ref{Tr}), the factor $e^{-\frac{R_{AV}^2}{R_0^2}}$ is introduced because the creation of quark-antiquark pairs should become less likely as the distance from the bare-hadron source is increased. $R_{AV}$ is the relative distance between the source particle and quark-antiquark pair in the vacuum. In Eq.~(\ref{T2}), there are three parameters need to identify, $\gamma, f$ and $R_0$.
According to our previous work~\cite{prd97094016}, we find
\begin{equation}\label{para}
\gamma = 32.2,\; \quad f=0.5\,\mbox{fm},\; \quad R_0=1\,\mbox{fm}.
\end{equation}
In the present work, we also apply the transition operator in Eq. (\ref{T2}) with improvements and remain the values of parameters $\gamma, f$ and $R_0$ in Eq. (\ref{para}). The newly mass shifts of the charmonium valence states are demonstrated in Table \ref{result2}.

From the table, we found that the mass shifts are reduced greatly by $75\%$ averagely, compared with the results in Table \ref{result1}. Plainly, our modified $^3P_0$ pair-creation model generates modest unquenching corrections, with mass renormalizations just $1\%-4\%$ of a given meson's bare mass. In Table \ref{mesonspectra}, we obtained the masses of the charmonium mesons in the quenched quark model, and the masses of $1S$, $2S$, $1P$, $2P$ and $1D$ are all smaller than the experimental values from PDG \cite{PDG}. In the unquenched quark model, the coupled-channel effects result in the negative mass shifts, which leads to the smaller unquenched masses for the states. Notably, although the mass shifts reported in Table \ref{result2} are sensible, they destroy agreement with the empirical masses. This is because the model parameters in Table \ref{modelparameters} were determined by fitting the meson spectrum from light to heavy, without considering the coupled-channel effects. As a exercise, we choose to illustrate a remedy. We adjust the confinement parameter $\Delta$ in order to increase the quenched masses for only $c\bar{c}$ mesons such that unquenching delivers the empirical masses, an outcome achieved with
\begin{equation}
%\alpha_0 = 3.85\,,\quad
\Delta = -62\,\mbox{MeV}.
\end{equation}
%引用赵光达的文献
The results are listed in Table \ref{result3}. We can find that the mass shifts are not very sensitive to the parameter $\Delta$. Having made our point, we leave for the future a complete refit of the parameters in Table~\ref{modelparameters} in order to arrive finally at a fully unquenched quark model.

%表格中的数据找规律，一个一个的分析每个态，Ds 是 D的1/3等等。
%质量劈裂问题
%对数据进行分析，和实验数据进行对比
Now let us focus on the numerical analysis on the results in Table \ref{result3}. Firstly, the mass shift of the each single coupled-channel $D_s^{(*)}D_s^{(*)}$ is smaller than, and is about $\frac{1}{3}$ of the $D^{(*)}D^{(*)}$, due to the $\gamma$ values. For open channels, the mass shifts are also larger than the close channels. For example, for $\chi_{c_0}(2P) \rightarrow D\bar{D}$, the mass shift is about 60 MeV. For $\chi_{c_1}(2P) \rightarrow D\bar{D}^* + D^*\bar{D}$ and  $h_c(2P) \rightarrow D\bar{D}^* + D^*\bar{D}$, the mass shift is about 58 MeV and 36 MeV, respectively, and they are all much larger than the other close channels. Because the coupling of all channels is not so significant, as an approximation, the ``Total" column represents the total mass shifts of the state, which is obtained by summing the mass shifts of the each coupled-channel simply.

Secondly, the $J/\psi-\eta_c$ and $\psi(2S)-\eta_c(2S)$ loop-induced mass splitting has been discussed previously by Eichten \emph{et al.} \cite{Eichten:2004uh}. The authors find a small loop-induced $J/\psi-\eta_c$ mass splitting of -3.7 MeV and a $\psi(2S)-\eta_c(2S)$ splitting of -20.9 MeV, bringing their model into good agreement with the experimental $\psi(2S)-\eta_c$ mass difference. Table \ref{result3} shows that we find a numerically similar $\psi(2S)-\eta_c(2S)$ splitting of -18.4 MeV, but the ground state $J/\psi-\eta_c$ mass difference is -13.4 MeV. In Ref. \cite{prc77055206}, Barnes and Swanson get the $\psi(2S)-\eta_c(2S)$ splitting of -24 MeV, which is well consistent with our results. But the $J/\psi-\eta_c$ mass difference is -34 MeV which is larger than ours.

Thirdly, let us compare our ``Unquenched mass'' (the last column in the table) with the experiment values. By simple correction of model confinement parameter $\Delta$ in Table \ref{modelparameters}, the bare masses of $c\bar{c}$ states are increased. After considering the coupled-channel effects, the unquenched masses of the states, $\eta_c(1S)$, $J/\psi$, $\chi_{c_0}(1P)$, $\chi_{c_1}(2P)$, $\chi_{c_2}(2P)$ and $h_c(2P)$ are well consistent with the experimental values.
In our future work, we will adjust the model parameters related to the charm quark in Table \ref{modelparameters} and keep the light meson sector unchanged as much as possible.

\begin{table*}[!t]
\begin{center}
\caption{ \label{result1} Mass shifts computed for $c\bar{c}$ charmonium mesons using the transition matrix constructed from $T_1$ in Eq. (\ref{Tr}) with $f \rightarrow 0$, $\gamma=6.95$. (Units of MeV)}
\begin{tabular}{cccccccccccccccc}
\hline \hline \multicolumn{3}{c}{Bare $c\bar{c}$ state}
&\multicolumn{10}{c}{Mass
shifts by channels} &\multicolumn{3}{c}{$c\bar{c}+qq\bar{q}\bar{q}$} \\
\cline{1-3}\cline{5-13}\cline{15-16}
 State$(n^{2S+1}L_J)$ &Bare mass &Exp &~~~ &$D\bar{D}$ &$D\bar{D^*}$ &$D^*\bar{D}$ &$D^*\bar{D^*}$ &$D_s\bar{D_s}$ &$D_s\bar{D}_s^*$ &$D^*_s\bar{D}_s$ &$D^*_s\bar{D}_s^*$ &Total & & &Unquenched mass \\
\hline
$\eta_c(1S)(1^1S_0)$         &2964.4 &2983.9             & &...    &-197.6 &-197.6 &-369.4 &...   &-80.7 &-80.7 &-155.8 &-1081.8 && &1882.6\\
$\eta_c(2S)(2^1S_0)$         &3507.8 &3637.5             & &...    &-127.8 &-127.8 &-228.7 &...   &-46.9 &-46.9 &-89.0  &-667.1  && &2840.6\\
$J/\psi(1S)(1^3S_1)$         &3096.4 &3096.0             & &-60.6  &-111.6 &-111.6 &-370.5 &-22.1 &-41.6 &-41.6 &-139.6 &-899.2  && &2197.2\\
$\psi(2S)(2^3S_1)$           &3605.0 &3686.1             & &-48.4  &-83.7  &-83.7  &-259.5 &-15.5 &-29.1 &-29.1 &-96.2  &-645.3  && &2959.7\\
$\chi_{c_0}(1P)(1^3P_0)$      &3362.8 &3414.7             & &-111.5 &...    &...    &-30.4  &-36.0 &...   &...   &-10.5  &-188.5  && &3174.3\\
$\chi_{c_1}(1P)(1^3P_1)$      &3393.9 &3510.7             & &...    &-65.3  &-65.3  &...    &...   &-20.7 &-20.7 &...    &-172.2  && &3221.7\\
$\chi_{c_2}(1P)(1^3P_2)$      &3435.8 &3556.2             & &...    &...    &...    &-113.8 &...   &...   &...   &-35.6  &-149.5  && &3286.3\\
$h_c(1P)(1^1P_1)$            &3416.1 &3525.4             & &...    &-32.5  &-32.5  &-57.3  &...   &-9.9  &-9.9  &-18.6  &-160.7  && &3255.4\\
$\chi_{c_0}(2P)(2^3P_0)$      &3814.7 &$\chi_{c_0}(3915)$?& &-130.8 &...    &...    &-29.8  &-33.7 &...   &...   &-9.8   &-204.1  && &3610.6\\
$\chi_{c_1}(2P)(2^3P_1)$      &3851.9 &$\chi_{c_1}(3872)$?& &...    &-74.2  &-74.2  &...    &...   &-19.8 &-19.8 &...    &-188.0  && &3663.9\\
$\chi_{c_2}(2P)(2^3P_2)$      &3901.1 &$\chi_{c_2}(3930)$?& &...    &...    &...    &-111.2 &...   &...   &...   &-34.8  &-145.9  && &3755.1\\
$h_c(2P)(2^1P_1)$            &3877.4 &$Z_c(3930)$?       & &...    &-49.8  &-49.8  &-57.8  &...   &-9.7  &-9.7  &-18.0  &-194.8  && &3682.6\\
$\eta_{c_2}(1D)(1^1D_2)$     &3675.1 &?                  & &...    &-32.2  &-32.2  &-53.8  &...   &-8.7  &-8.7  &-16.2  &-151.8  && &3523.4\\
$\psi(1D)(1^3D_1)$           &3653.3 &$\psi(3770)$?      & &-62.8  &-27.0  &-27.0  &-18.5  &-16.3 &-7.5  &-7.5  &-5.6   &-172.1  && &3481.2\\
$\psi_2(1D)(1^3D_2)$         &3668.3 &$\psi_2(3823)$?    & &...    &-47.4  &-47.4  &-28.5  &...   &-13.1 &-13.1 &-8.2   &-157.8  && &3510.5\\
$\psi_3(1D)(1^3D_3)$         &3688.1 &$\psi_3(3842)$?    & &...    &...    &...    &-106.5 &...   &...   &...   &-31.6  &-138.2  && &3549.9\\
 \hline \hline
\end{tabular}
\end{center}
\end{table*}

\begin{table*}[!t]
\begin{center}
\caption{ \label{result2} Mass shifts computed for $c\bar{c}$ charmonium mesons using the transition matrix constructed from $T_2$ in Eq. (\ref{T2}) with $f=0.5$ fm, $\gamma=32.2$, $R_0=1$ fm. (Units of MeV)}
\begin{tabular}{cccccccccccccccc}
\hline \hline \multicolumn{3}{c}{Bare $c\bar{c}$ state}
&\multicolumn{10}{c}{Mass
shifts by channels} &\multicolumn{3}{c}{$c\bar{c}+qq\bar{q}\bar{q}$} \\
\cline{1-3}\cline{5-13}\cline{15-16}
 State$(n^{2S+1}L_J)$ &Bare mass &Exp &~~~ &$D\bar{D}$ &$D\bar{D^*}$ &$D^*\bar{D}$ &$D^*\bar{D^*}$ &$D_s\bar{D_s}$ &$D_s\bar{D}_s^*$ &$D^*_s\bar{D}_s$ &$D^*_s\bar{D}_s^*$ &Total & & &Unquenched mass \\
\hline
$\eta_c(1S)(1 ^1S_0)$        &2964.4 &2983.9             & &...    &-14.5  &-14.5  &-27.1  &...   &-3.2  &-3.2   &-6.4   &-68.9   && &2895.4\\
$\eta_c(2S)(2^1S_0)$         &3507.8 &3637.5             & &...    &-31.6  &-31.6  &-53.1  &...   &-4.9  &-4.9   &-9.1   &-135.3  && &3372.5\\
$J/\psi(1S)(1^3S_1)$         &3096.4 &3096.0             & &-6.4   &-11.8  &-11.8  &-38.9  &-1.4  &-2.6  &-2.6   &-8.7   &-84.2   && &3012.2\\
$\psi(2S)(2^3S_1)$           &3605.0 &3686.1             & &-16.5  &-25.2  &-25.2  &-71.2  &-2.1  &-3.7  &-3.7   &-11.9  &-159.5  && &3445.5\\
$\chi_{c_0}(1P)(1^3P_0)$      &3362.8 &3414.7             & &-20.5  &...    &...    &-5.2   &-3.4  &...   &...    &-1.0   &-30.1   && &3332.7\\
$\chi_{c_1}(1P)(1^3P_1)$      &3393.9 &3510.7             & &...    &-12.9  &-12.9  &...    &...   &-2.2  &-2.2   &...    &-30.3   && &3363.5\\
$\chi_{c_2}(1P)(1^3P_2)$      &3435.8 &3556.2             & &...    &...    &...    &-25.1  &...   &...   &...    &-4.5   &-29.6   && &3406.1\\
$h_c(1P)(1^1P_1)$            &3416.1 &3525.4             & &...    &-6.9   &-6.9   &-11.7  &...   &-1.2  &-1.2   &-2.2   &-30.1   && &3386.0\\
$\chi_{c_0}(2P)(2^3P_0)$      &3814.7 &$\chi_{c_0}(3915)$?& &-106.5 &...    &...    &-11.7  &-6.6  &...   &...    &-1.3   &-126.1  && &3688.6\\
$\chi_{c_1}(2P)(2^3P_1)$      &3851.9 &$\chi_{c_1}(3872)$?& &...    &-49.7  &-49.7  &...    &...   &-3.4  &-3.4   &...    &-106.2  && &3745.7\\
$\chi_{c_2}(2P)(2^3P_2)$      &3901.1 &$\chi_{c_2}(3930)$?& &...    &...    &...    &-56.3  &...   &...   &...    &-6.0   &-62.3   && &3838.8\\
$h_c(2P)(2^1P_1)$            &3877.4 &$Z_c(3930)$?       & &...    &-44.8  &-44.8  &-28.2  &...   &-1.8  &-1.8   &-2.9   &-124.3  && &3753.1\\
$\eta_{c_2}(1D)(1^1D_2)$     &3675.1 &?                  & &...    &-12.1  &-12.1  &-18.3  &...   &-1.6  &-1.6   &-2.8   &-48.4   && &3626.8\\
$\psi(1D)(1^3D_1)$           &3653.3 &$\psi(3770)$?      & &-25.4  &-9.3   &-9.3   &-5.8   &-2.8  &-1.2  &-1.2   &-0.9   &-55.9   && &3597.3\\
$\psi_2(1D)(1^3D_2)$         &3668.3 &$\psi_2(3823)$?    & &...    &-17.4  &-17.4  &-9.5   &...   &-2.3  &-2.3   &-1.4   &-50.2   && &3618.0\\
$\psi_3(1D)(1^3D_3)$         &3688.1 &$\psi_3(3842)$?    & &...    &...    &...    &-38.3  &...   &...   &...    &-5.8   &-44.1   && &3644.1\\
 \hline \hline
\end{tabular}
\end{center}
\end{table*}

\begin{table*}[!t]
\begin{center}
\caption{ \label{result3} Mass shifts computed for $c\bar{c}$ charmonium mesons using the transition matrix constructed from $T_2$ in Eq. (\ref{T2}) with $f=0.5$ fm, $\gamma=32.2$, $R_0=1$ fm and $\Delta=62$ MeV. (Units of MeV)}
\begin{tabular}{cccccccccccccccc}
\hline \hline \multicolumn{3}{c}{Bare $c\bar{c}$ state}
&\multicolumn{10}{c}{Mass
shifts by channels} &\multicolumn{3}{c}{$c\bar{c}+qq\bar{q}\bar{q}$} \\
\cline{1-3}\cline{5-13}\cline{15-16}
 State$(n^{2S+1}L_J)$ &Bare mass &Exp &~~~ &$D\bar{D}$ &$D\bar{D^*}$ &$D^*\bar{D}$ &$D^*\bar{D^*}$ &$D_s\bar{D_s}$ &$D_s\bar{D}_s^*$ &$D^*_s\bar{D}_s$ &$D^*_s\bar{D}_s^*$ &Total & & &Unquenched mass \\
\hline
$\eta_c(1S)(1 ^1S_0)$        &3051.3 &2983.9             & &...    &-13.7  &-13.7  &-25.7  &...   &-3.1  &-3.1   &-6.1  &-65.4    && &2985.9\\
$\eta_c(2S)(2^1S_0)$         &3594.7 &3637.5             & &...    &-27.9  &-27.9  &-48.2  &...   &-4.5  &-4.5   &-8.4  &-121.3   && &3473.4\\
$J/\psi(1S)(1^3S_1)$         &3183.3 &3096.0             & &-5.9   &-11.0  &-11.0  &-36.3  &-1.3  &-2.4  &-2.4   &-8.3  &-78.8    && &3104.6\\
$\psi(2S)(2^3S_1)$           &3691.9 &3686.1             & &-13.2  &-21.6  &-21.6  &-63.9  &-1.8  &-3.4  &-3.4   &-10.9 &-139.7   && &3552.3\\
$\chi_{c_0}(1P)(1^3P_0)$      &3449.7 &3414.7             & &-18.1  &...    &...    &-4.7   &-3.1  &...   &...    &-1.0  &-26.9    && &3422.8\\
$\chi_{c_1}(1P)(1^3P_1)$      &3480.8 &3510.7             & &...    &-11.5  &-11.5  &...    &...   &-2.1  &-2.1   &...   &-27.2    && &3453.5\\
$\chi_{c_2}(1P)(1^3P_2)$      &3522.7 &3556.2             & &...    &...    &...    &-22.7  &...   &...   &...    &-4.2  &-26.9    && &3495.8\\
$h_c(1P)(1^1P_1)$            &3502.9 &3525.4              & &...    &-6.1   &-6.1   &-10.7  &...   &-1.1  &-1.1   &-2.0  &-27.1    && &3475.9\\
$\chi_{c_0}(2P)(2^3P_0)$      &3901.7 &$\chi_{c_0}(3915)$? & &-60.5  &...    &...    &-9.2   &-4.8  &...   &...    &-1.2  &-75.8    && &3825.9\\
$\chi_{c_1}(2P)(2^3P_1)$      &3938.8 &$\chi_{c_1}(3872)$?& &...    &-28.6  &-28.6  &...    &...   &-2.8  &-2.8   &...   &-62.9    && &3875.9\\
$\chi_{c_2}(2P)(2^3P_2)$      &3988.0 &$\chi_{c_2}(3930)$?& &...    &...    &...    &-42.1  &...   &...   &...    &-5.2  &-47.3    && &3940.7\\
$h_c(2P)(2^1P_1)$            &3964.3 &$Z_c(3930)$?       & &...    &-17.8  &-17.8  &-20.7  &...   &-1.5  &-1.5   &-2.5  &-61.8    && &3902.5\\
$\eta_{c_2}(1D)(1^1D_2)$     &3762.1 &?                  & &...    &-9.8   &-9.8   &-15.9  &...   &-1.4  &-1.4   &-2.5  &-41.1    && &3721.0\\
$\psi(1D)(1^3D_1)$           &3740.2 &$\psi(3770)$?      & &-19.2  &-7.6   &-7.6   &-5.1   &-2.5  &-1.1  &-1.1   &-0.8  &-45.3    && &3694.9\\
$\psi_2(1D)(1^3D_2)$         &3755.2 &$\psi_2(3823)$?    & &...    &-14.3  &-14.3  &-8.2   &...   &-2.1  &-2.1   &-1.3  &-42.3    && &3712.8\\
$\psi_3(1D)(1^3D_3)$         &3775.0 &$\psi_3(3842)$?    & &...    &...    &...    &-33.3  &...   &...   &...    &-5.3  &-38.6    && &3736.4\\
 \hline \hline
\end{tabular}
\end{center}
\end{table*}

\begin{table}[!t]
\begin{center}
\caption{ \label{compare} The comparison of the mass shifts of the charmonium mesons between several theoretical works. The $2^{nd}$ and $3^{rd}$ column denotes our results corresponding to Table \ref{result1} and Table \ref{result2}. (Units of MeV)}
\begin{tabular}{cccccccc}
\hline \hline
States       &$\Delta M_1$ &$\Delta M_2$  &\cite{prd80014012} &\cite{prd76077502}&\cite{Monteiro:2018rkg} &\cite{prd72034010} &\cite{prc77055206}  \\
\hline
$1^1S_0$   &-1081.8   &-68.9   &-148    &-148   &-208  &-165    &-423                                                                                    \\
$2^1S_0$   &-667.1    &-135.3  &-208    &-158  &-84   &-200    &-416                                                                                           \\
$1^3S_1$   &-899.2    &-84.2   &-159    &-148   &-238  &-177    &-457                                                                                       \\
$2^3S_1$   &-645.3    &-159.5  &-228    &-157   &-99   &-216    &-440                                                                                        \\
$1^3P_0$   &-188.5    &-30.1   &-181    &-157   &-141  &-198    &-459                                                                                         \\
$1^3P_1$   &-172.2    &-30.3   &-195    &-173   &-191  &-215    &-496                                                                                           \\
$1^3P_2$   &-149.5    &-29.6   &-210    &-154   &-218  &-228    &-521                                                                                         \\
$1^1P_1$   &-160.7    &-30.1   &-201    &-150   &-189  &-219    &-504                                                                                           \\
$2^3P_0$   &-107.3    &-34.4   &-179    &-218   &-38   &...     &...                                                                                       \\
$2^3P_1$   &-52.1     &-18.8   &-300    &-214   &-58   &...     &...                                                                                            \\
$2^3P_2$   &-145.9    &-62.3   &-268    &-203   &-64   &...     &...                                                                                            \\
$2^1P_1$   &-127.2    &-59.7   &-230    &-153   &-51   &...     &...                                                                                           \\
$1^1D_2$   &-151.8    &-48.4   &-226    &...    &-112  &...     &...                                                                                            \\
$1^3D_1$   &-172.1    &-55.9   &-233    &-188   &-125  &...     &...                                                                                             \\
$1^3D_2$   &-157.8    &-50.2   &-226    &...    &-121  &...     &...                                                                                           \\
$1^3D_3$   &-138.2    &-44.1   &-230    &...    &-116  &...     &...                                                                                            \\ \hline \hline
\end{tabular}
\end{center}
\end{table}

What's more, for comparisons, we show some theoretical results about the mass shifts of charmonium mesons in Table \ref{compare}. In the table, the mass shifts have minus sign overall. The second column $\Delta M_1$ are the mass shifts with the original transition operator of the $^3P_0$ model. For $1S$ and $2S$ states, the mass shifts are larger than the other theoretical works in Table \ref{compare}. For $1P$ states, our results are basically consistent with the references \cite{prd76077502,prd72034010,prd80014012,Monteiro:2018rkg}. Therein, the spherical harmonic oscillator (SHO) wave function is applied to describe the meson dynamics, and the relative motion between two mesons is described by plane-wave functions. Besides, the mass shifts are dependent on the parameter $\beta$ in SHO and $\gamma$ in the $^3P_0$ model. The systematic errors due to the approximations are unpredictable for the bound-state calculation, although they are not a bad approximation for the decay width calculation. Our results $\Delta M_2$ in the third column, obtained with the improved $^3P_0$ model, are comparable to each other, but much smaller than the other theoretical results in 4-8 columns.

%What is more important is not the comparison of the mass shifts simply, but is the obtained unquenched mass of the state by readjusting of the model parameters with the help of the current experimental data.

%%%%%%%%%%%%%%%%%%%%%%%%%%%%%%%%%%%%%%%%%%%%%%%%%%%%%%%%%%%%%

\section{Summary}
\label{epilogue}
In the present work, the spectrum of $1S$, $2S$, $1P$, $2P$ and $1D$ charmonium states below 4 GeV is calculated taking into account coupling to the pairs of lowest $D$ and $D_s$ pairs. To minimize the error from the calculation, a powerful method for dealing with few-body systems (GEM) was used. In our work, the angular momentum of the two mesons takes zero, and the relative motion between the two mesons denotes to $P$ wave for $1S$, $2S$ and $1D$ states. For $1P$ and $2P$ states, we only consider the relative motion to be $S$ wave for the preliminary work, and $D$ wave related calculations will be our future work.

The transition operator of the $^3P_0$ model is required to relate the valence part to the four-quark components. We demonstrated the mass shifts of the charmonium states with the original transition operator of the $^3P_0$ model, as well as with the modified version of the transition operator. By contrast, the masses shifts are reduced greatly by $75\%$ averagely within the modified $^3P_0$ model. Plainly, our modified $^3P_0$ pair-creation model generates modest unquenching corrections, with mass renormalizations just $1\%-4\%$ of a given meson’s bare mass.

As a preliminary work, we only fine-tune the model confinement parameter $\Delta$, and we obtained unquenched masses for the charmonium states. We find that the masses of the charmonium states, $\eta_c(1S)$, $J/\psi$, $\chi_{c_0}(1P)$, $\chi_{c_1}(2P)$, $\chi_{c_2}(2P)$ and $h_c(2P)$ are well consistent with the experimental values. We leave for the future a complete refit of the model parameters in order to arrive finally at a fully unquenched quark model. More experimental data in the future can help us better understand the spectrum of the charmonium states.

\acknowledgments
This work is partly supported by the National Natural Science Foundation of China under Grants No. 12205125, No. 11847145, No. 12205249 and No. 11865019, and also supported by the Natural Science Foundation of Jiangsu Province under Grants No. BK20221166.

%%%%%%%%%%%%%%%%%%%%%%%%%%%%%%%%%%%%%%%%%%%%%%%%%%%%%%%%%%%%%%%%%%%%%%%%%%%%%%%%%%%%%%%%%%%%%%%%%%%%%%%%%%%%%%%%%%%%%%%

\end{document}